\documentclass[10pt,letterpaper]{article}

\usepackage{cogsci}

\cogscifinalcopy %
\usepackage[natbibapa]{apacite}

\usepackage{amssymb}
\usepackage{amsmath}
\usepackage{siunitx}

\usepackage[hidelinks]{hyperref}

\usepackage{graphicx}
\usepackage{pslatex}
\usepackage{apacite}
\usepackage{float} %

\title{Harmonizing Program Induction with Rate-Distortion Theory}
 
\author{\textbf{Hanqi Zhou$^{1,2,3,4}$}(hanqi.zhou@uni-tuebingen.de), \textbf{David G. Nagy$^{1,2,4}$} \& \textbf{Charley M. Wu$^{1,2,3,4}$} \\
$^1$ Human and Machine Cognition Lab, University of T\"ubingen, T\"ubingen, Germany\\
$^2$ Max Planck Institute for Biological Cybernetics $^3$ IMPRS-IS $^4$ T\"ubingen AI Center\\
}

\begin{document}

\maketitle

\begin{abstract}
Many aspects of human learning have been proposed as a process of constructing mental programs: from acquiring symbolic number representations to intuitive theories about the world. In parallel, there is a long-tradition of using information processing to model human cognition through Rate Distortion Theory (RDT). 
Yet, it is still poorly understood how to apply RDT when mental representations take the form of programs. 
In this work, we adapt RDT by proposing a three way trade-off among rate (description length), distortion (error), and computational costs (search budget).
We use simulations on a melody task to study the implications of this trade-off, and show that constructing a shared program library across tasks provides global benefits. However, this comes at the cost of sensitivity to curricula, which is also characteristic of human learners. Finally, we use methods from partial information decomposition to generate training curricula that induce more effective libraries and better generalization.

\end{abstract}

\section{Introduction}
Human learning is not just the accumulation of knowledge, but can be seen as a process of distilling experience into compact, generalizable programs \citep{rule2020child}. This process requires balancing the richness of our mental representations with the cognitive cost of maintaining and processing them \citep{lieder2020resource}. %
For example, when learning music, people typically start by learning to reproduce specific melodies verbatim, through which we develop a general understanding of musical structures. Simultaneously, we need to develop compact representations, with seasoned musicians being able to utilize regularities such as key or scale to more efficiently encode melodies without needing to memorize each note. These dual aspects of program learning and compression are crucial for efficient recognition, recall, and creation of new melodies. %

\emph{Program induction} is the process of inferring rules or instructions that generate an observed pattern of data \citep{rule2020child, ellis2021dreamcoder}. Emerging from early computational theories \citep{solomonoff1964formal, gold1967language} formalizing learning as an algorithmic process of inferring the program that generated the observed data, it has since been refined by cognitive scientists using symbolic reasoning and logical programming as a framework to model human learning and thinking \citep{fodor1975language}. This approach offers a model for understanding how humans, or machines, learn underlying rules, structures, or more broadly, generative program-like representations from observed examples.

\emph{Rate distortion theory} \citep[RDT;][]{shannon1959coding, berger1971rate} provides a normative framework for how to allocate limited resources to encode information in settings where perfect (i.e., lossless) compression is not possible. Since humans have limited cognitive resources, RDT has provided a powerful framework for modeling how we develop informative and structured mental representations, selectively ignoring irrelevant or redundant information to most effectively encode the most useful and relevant information. %
Recently, RDT has been applied to a wide number of cognitive phenomena, such as memory \citep{nagy2020optimal, gershman2021rational, bates2020efficient}, perception \citep{sims2016rate, bates2019adaptive}, and decision-making \citep{bhui2018decision, lai2021policy}.

\begin{figure}
\centering
  \includegraphics[width=\linewidth]{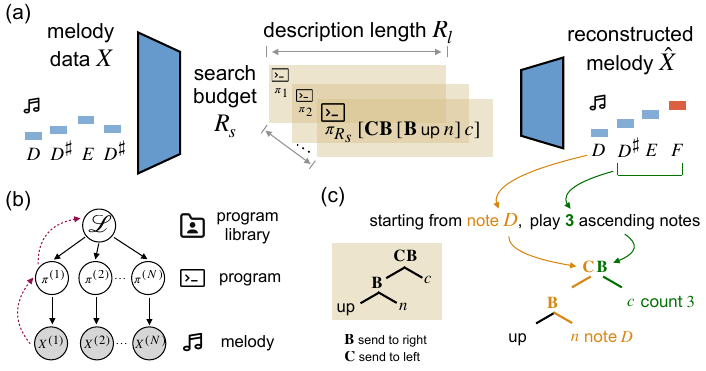}
    \vspace{-1em}
  \caption{(a) Program induction under resource constraints using an encoder-decoder framework on melody data. The encoder compresses melodies (piano rolls~$X$) onto a latent space (programs~$\pi$), which are constrained by two bottlenecks: description length $R_l$ and search budget $R_s$.
  (b) Each melody $X^{(i)}$ is assumed to be generated by program $\pi^{(i)}$, which is defined by a program library~$\mathcal{L}$ (solid arrows). When inferring the program and library given observed melodies (dashed arrows), the goal is to find a balance between compact and easy-to-search programs, while minimizing reconstruction error~$d(X,\hat{X})$. (c) Illustrative example of the tree structure and routers used in a program. }
  \label{fig:framework}
\end{figure}

However, it is not well understood how to apply the RDT framework when mental representations take the form of programs, and what limitations might arise. In particular, program induction implies path dependencies arising from the iterative construction of a library of programs, and there are open questions concerning how the curriculum order of learning materials influences the ease of generalizing to new data.

\subsection{Goal and scope}
Here, we model program induction under resource constraints with RDT in the domain of learning musical melodies. Through simulations, we show that two different dimensions of compression are relevant to learning using adaptor grammars \citep[AGs;][]{johnson2006adaptor}. On the one hand and consistent with past algorithmic RDT approaches, there is a bias towards learning simpler programs \cite{vereshchagin_algorithmic_2005,sow_complexity_2003,sable-meyer_language_2022}.
On the other hand, the amount of cognitive resources available to search for new programs adds a new dimension that is not accounted for in RDT.
Together, these two dimensions influence path-dependent learning, where we show that building a more synergistic library of programs with greater overlap between stimuli allows for better generalization. 
Altogether, this work provides the first steps towards integrating program induction with RDT, showing new dimensions of relevance and providing guidance for future work applying this framework for modeling human learners.

\section{Methods} 
In this paper, we study the interaction between compression and program learning using simulations on a melody learning task (Fig.~\ref{fig:framework}). The model is presented with successive notes from a melody, which needs to be encoded to minimize distortion (i.e., error), while subject to limitations on the description length of the program (i.e., memory) and resources available for searching over programs (i.e., computational costs).

\subsection{Melody learning task}
Music is a universal element of human culture, transcending linguistic and cultural boundaries \citep{mehr2019universality, loui2010humans, jacoby2024commonality}. The structured and inventive nature of music makes it perfect for studying a compositional domain where simpler elements (notes, rhythms) can be combined into more complex structures (melodies, harmonies).
Our dataset comes from real-world melodies in the form of monophonic piano rolls \citep[Fig.~\ref{fig:framework}a;][]{garciavalencia2020sequence}.

We first pre-process the piano rolls by 1) removing melodies with fewer than 10 notes, 2) normalizing notes into a single-octave range with 12 notes [$C, D, E, F, G, A, B, C, C\#, D\#, F\#, G\#, A\#$], 
and 3) denoting pauses with a special symbol to exclude them from arithmetic operations. 
After pre-processing, we randomly sampled 1,000 melodies for a training set and 500 melodies for an evaluation set, with mean lengths of 50 notes in each. 
We then train probabilistic encoder-decoder models (Fig.~\ref{fig:framework}a) to compress melodies using program-like representations and test how they generalize to new melodies.
The model encodes each melody onto a latent space, by searching over discrete programs, subject to constraints on memory and computational costs.
We first introduce a general Bayesian program induction framework and provide details about these constraints.

\subsection{Bayesian program induction}
\textit{Bayesian program induction} \citep{goodman2008rational, piantadosi2012bootstrapping, lake2015human, rule2018learning, correa2023exploring} provides a modern interpretation of Fodor’s \citeyear{fodor1975language} \textit{Language of Thought} (LoT), allowing for both probabilistic representations of stimuli and the ability to simulate new scenarios given a set of learned programs. In our melody task, the goal is to find a program~$\pi$ that adequately represents each melody $X^{(i)} \in \{X^{(1)},\ldots X^{(N)}\}$, described by a sequence of $T$ notes $X^{(i)} = [x_1^{(i)}, \ldots, x_T^{(i)}]$. By specifying a prior probability distribution over all programs~$p(\pi)$, and the likelihood of the observed melody under each program~$p(X \,|\, \pi)$, Bayes’ theorem can be used to compute the posterior probability of the encoding program conditioned on the observed melody:
\begin{align}
 p(\pi \,|\, X) \propto p(X \,|\, \pi) p(\pi)
\end{align}

Here, we use \textit{combinatory logic} \citep[CL;][]{liang2010learning,zhao2023model, schonfinkel_uber_1924}, which, although predating it, can be seen as a variant of $\lambda-$calculus \citep{piantadosi2012bootstrapping, ellis2021dreamcoder, church1936unsolvable} that eliminates the need to keep track of variable names and scopes. This crucial feature helps identify shared substructures between programs \citep{liang2010learning}. 
A program~$\pi$ in CL can be represented as a binary tree, where for example, the program for ascending by three semitones can be described as $\pi = [\mathbf{CB},[\mathbf{B},\texttt{up}, n], c]$ (Fig.~\ref{fig:framework}c). There are two arguments with different \emph{types}, where the \emph{router}~$\mathbf{CB}$ directs the first argument~$n$ to the left subtree~$[\mathbf{B},\texttt{up}, n]$ (orange branch in Fig.~\ref{fig:framework}c), and the second argument~$c$ to the right subtree (green branch in Fig.~\ref{fig:framework}c). The \emph{primitive} $\texttt{up}$ in the left subtree defines the operation on the provided arguments (e.g., note 2 and count 3). We now provide a brief overview of the various components; however due to space constraints, for a more comprehensive explanation see \citet{liang2010learning}.

\paragraph{Routers.} %
Whereas CL originally required only two elementary combinators ($\mathbf{S}$ and $\mathbf{K}$) to sufficiently instantiate any computable function,  \citet{liang2010learning} introduced higher-order combinators called \textit{routers}~$\mathcal{R}$ to make computations more practical. 
Routers~$\mathcal{R}$ are finite sequences composed of combinators (first-order routers) $\{\mathbf{B},\mathbf{C},\mathbf{S}\}$ and direct incoming variables to the corresponding subtrees. 
For instance, in the generic program [$\mathcal{R}, \Gamma_l, \Gamma_r$], 
the router~$\mathbf{B}$ first directs the variable to the right subtree ~$\Gamma_r$ with the results then sent to $\Gamma_l$;
the router~$\mathbf{C}$ directs the variable first to the left tree; and router~$\mathbf{S}$ sends the variable to both trees. 
While abstract, these routers can compose complex functions from simpler ones, enabling infinitely productive program generation by combining a small set of primitives (base and function terms).
 
\paragraph{Primitives.}
Primitives, consisting of \emph{base terms} and \emph{function terms}, are both pre-defined. 
In our melody task, we define \emph{base terms} as note~$n$, count~$c$, and time~$m$ by assuming learners have an initial library~$\mathcal{L}$ over simple integers. 
\emph{Function terms} are interpreted as functions that can take typed arguments as input. For example, primitive~$\texttt{rep}(n, c)$ takes two arguments, note~$n$ and count~$c$, representing repeating note~$n$ for $c$ times (e.g., $\texttt{rep}([C], 2) = [C,C]$). Another example, $\texttt{get}(n, m)$, takes two arguments, note~$n$ and time index~$m$, returning the first~$m$ notes in note array~$n$ (e.g., $\texttt{get}([C,D,D^\#], 2) = [C,D]$). 
Both \emph{base terms} and \emph{function terms} can work as inputs to other function terms, e.g., $\texttt{get}(\texttt{rep}(n, c), m)$. 

\paragraph{Types.} 
Typed CL assigns different types~$t$ to the domain (i.e., the set of possible inputs) and codomain (i.e., the set of possible outputs) of primitives and programs, so that the program's syntactical correctness can be ensured. 
\emph{Base terms} (notes, counts, and times) are represented as distinct types $t_n, t_c, t_m$. This implies that, despite having identical numerical values (e.g., piano note 1, repeat 1 time, the first time index in an array), different integers have different physical meanings due to their distinct types. 
The type of a \emph{function term} is denoted using a right arrow $\rightarrow$. For instance, the type for the function term~$\texttt{get}(n, m)$ is expressed as $t_n, t_m \rightarrow t_n$, indicating the types of its input arguments and the type of its output.

\paragraph{Prior over programs.}
To define a prior over the tree representation of CL programs, we follow \citet{liang2010learning} and introduce a \textit{probabilistic context-free grammar} (PCFG). Intuitively, the nodes of the tree are successively expanded into subtrees according to pre-defined rules, and the complexity of the program is connected to the number of expansions, as well as the likelihoods of the terms. 
In order to generate samples from the PCFG while also maintaining syntactical correctness through types, we follow the generative process~$\operatorname{GENINDEP}(t)$ in \cite{liang2010learning} for each type~$t$. When sampling primitives and routers, the prior over different primitives and routers are equiprobable conditioned on types. Thus, the prior probability of a program $\pi = [\mathrm{r}, x, y]$ given type~$t$ is 
\begin{align} \label{eq:prog-prior}
    \log p(\pi \,|\, t) = \log p(x \,|\, t) + \log p(y \,|\, t) + \log p(\mathrm{r}).
\end{align}
There are two kinds of possible expansions. The first method involves sampling a router (e.g., $\mathbf{B}$) and a primitive (e.g., $\texttt{up}(n)$) that meets the type requirement based on prior $p(f \,|\, t)$ and $p(\mathrm{r})$, resulting in a program $[\mathbf{B}, \texttt{up}, n ]$. The second method recursively generates $K$ intermediate types $t_i$ along with their respective primitives and routers, ensuring the type sequence ultimately conforms to $t_n \rightarrow t_{i1} \rightarrow \ldots \rightarrow t_{iK} \rightarrow t_n$, thus yielding more nested programs.

\subsection{Compression with programs} 
Here, we are interested in how program induction is shaped by constraints on cognitive resources. To formalize this question, we rely on \textit{rate distortion theory} (RDT), which has been proposed as a normative framework for incorporating memory and processing constraints for biological agents \citep{sims2016rate, bates2020efficient, nagy2020optimal}. RDT characterizes the balance between minimizing \textit{distortion}~$D$ (the accuracy of the representation), while needing to keep the \textit{rate}~$R$ (the required resource) below a certain capacity limit. Formally, RDT typically defines rate as the mutual information~$I(X,Z)$ between the input~$X$ and its encoding~$Z$, and minimizes the distortion subject to a constraint~$R$ on this quantity:
\begin{align}
    D(R)=\text{inf}_{Q} \; D_Q, \;\; s.t.\; I(X,Z)\leq R,
\end{align}
where $Q$ represents the encoding function that defines the mapping~$X \mapsto \hat{X}$, and $D=\mathbb{E}[d(X,\hat{X})]$ is the expected distortion between the original observation~$X$ and the reconstruction~$\hat{X}$. Equivalently, the rate can be minimized such that the distortion is kept below the threshold $D$, and the minimal achievable rate for each constraint value defines the rate distortion function $R(D)=\inf_{Q} R_Q \;\; s.t.\; \mathbb{E}[d(X,\hat{X})]\leq D$.

\paragraph{Description length.} 
Here, we adapt RDT for programs by measuring rate using the description length of the program \citep{vereshchagin_algorithmic_2005, sow_complexity_2003,sable-meyer_language_2022}, which can be seen as a generalization of Kolmogorov complexity $K$ for imperfect reconstruction. This approach aligns with past theories suggesting human cognition has a preference for simpler representations \citep{chater2003simplicity, rubino2023compositionality, feldman2000minimization} and allows us to define RD functions for individual observations: 
\begin{align}
R_l(D)=\min_Q K(\hat{X}^{(i)}) \;\; s.t.\;  d(X^{(i)},\hat{X}^{(i)}) \leq D. \label{eq:algorithmicRDT}
\end{align}
Under certain assumptions, these functions asymptotically correspond to the Shannon RD curves \citep{vereshchagin_algorithmic_2005}. We assume that encodings of primitives are chosen optimally, and therefore measure the description length of a program $\pi_j$ as its negative log probability according to the prior over programs $l(\pi)=-\log p(\pi)$ \citep{grunwald2007minimum}. For the distortion function, following \citet{vereshchagin_algorithmic_2005}, we use Hamming distance, which counts how many of the recalled notes differ from the observed ones.

\paragraph{Search budget.}  Program induction, together with the optimization of the algorithmic RD objective (Eq.~\ref{eq:algorithmicRDT}), implies a search over the space of possible programs, necessitating the use of approximations. We model constraints on this search process by defining an upper limit on how many programs can be considered, which we refer to as the search budget~$R_s$. Extending Eq.~\ref{eq:algorithmicRDT} with this additional constraint provides a computationally bounded version of the RD objective:
\begin{align}
D_\text{bounded}(R)=\inf_{Q} \; D_Q, \;\; s.t.\; K(\hat{X}^{(i)})\leq R_l \; \land \; N_s\leq R_s, \label{eq:RDTwoBounds}
\end{align}
where $N_s$ is the number of considered programs. 

\subsection{Approximate inference}
Having introduced both Bayesian program induction and a computationally bounded version of algorithmic RDT, we now connect the two frameworks through approximate inference. Intuitively, both frameworks imply searching for programs that can compactly represent the observed melodies. In Bayesian inference, this is encouraged by the PCFG prior (Eq.~\ref{eq:prog-prior}), which attaches higher probabilities to shorter programs, whereas Eq.~\ref{eq:algorithmicRDT} imposes a direct limit on program length. To see how the search budget can be incorporated into the inference, we approximate the posterior using Monte Carlo (MC) methods via importance sampling. Here, proposed programs are sampled according to the prior $p(\pi)$, and then weighted by their likelihoods $p(X|\pi)$. This latter step corresponds to evaluating how likely each proposed program is to have generated observed melody, which from the RDT view corresponds to evaluating the distortion function on the reconstruction. Similar to other MC algorithms, the computational budget can be varied through the number of sampled proposals, which we identify with $N_s$ in Eq.~\ref{eq:RDTwoBounds}.

If the search over programs fails to find a sufficiently short candidate, we delete terms from the program in order of decreasing likelihood until the program satisfies the length constraint~$R_l$, and then sample missing terms from the prior during reconstruction. We interpret this process as analogous to a human learner constructing an overly complex program and forgetting parts of it, resulting in characteristic, structured noise. For instance, for $\pi=[\mathbf{CB},[\mathbf{B},\texttt{up}, n_2], c_3]$ (Fig.~\ref{fig:framework}a), the subject's reconstruction might ascend by four notes (instead of three). 
We introduce one additional approximation to make inference more tractable, where similar to how humans might listen to music, each melody can be segmented into subsequences represented by subprograms. %
Specifically, the program~$\pi^{(i)}$ for each melody $X^{(i)}$ can be composed of~$M^{(i)} \in \mathbb{N}^+$ subprograms~$\pi^{(i)} = \{\pi_1, \ldots, \pi_{M} \}$. 
Each subprogram~$\pi_j \in \pi^{(i)} \subseteq \Pi$ encodes a subsequence of the melody $X_{t_j:t_j'}^{(i)}=[x^{(i)}_j, \ldots, x^{(i)}_{j'}]$, with the length determined by the subprogram itself. We then evaluate proposals for these subprograms successively based on the likelihood.

\section{Results} 
We first extend the RDT framework to a program induction setting, where we examine the constraints of description length and search budget. Then we explore how \textit{adaptor grammars} (AGs) provide a method to enhance computational efficiency within these constraints. However, AGs are sensitive to order effects, resulting in large variations in generalization. Therefore, we develop a method to generate \textit{a priori} curricula that lead to improved generalization through an information-theoretic definition of synergy.

\begin{figure}[t]
\centering
  \includegraphics[height=6.2cm]{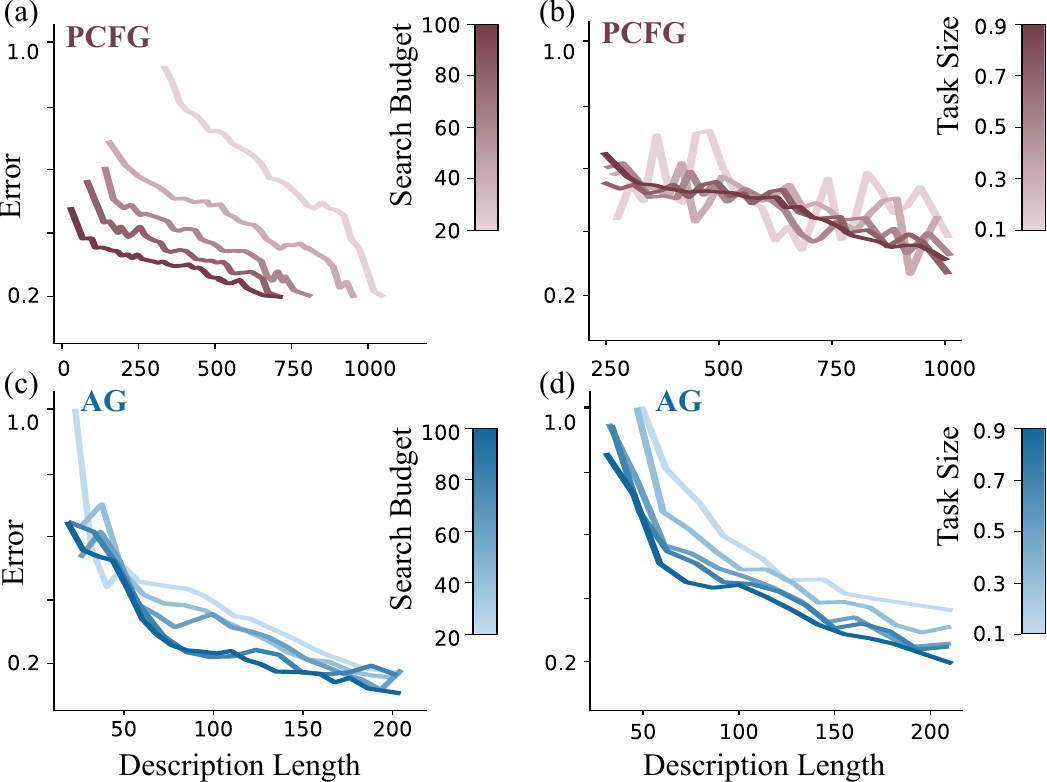}
  \caption{RD curves with (a-b) PCFGs and (c-d) AGs, given different description lengths and search budgets (left) and under different amounts of training data (right). }
  \label{fig:rdt-result}
  \vspace{-1em}
\end{figure}

To demonstrate how program induction is affected by the constraints of program length~$R_l$ and search budget~$R_s$, we first compute the RD curve using a PCFG model for a range of settings (Fig.~\ref{fig:rdt-result}a). These curves show that lower distortion is achieved through both longer programs and larger search budgets. 
However, one limitation of using a PCFG is that it relies on a set of pre-specified rules and primitives, without a method for adapting them to new observations, treating each new input as a separate task. %
In contrast, human learning is cumulative and contextual, allowing us to build upon what we have learned from past experiences.
Figure~\ref{fig:rdt-result}b illustrates how the RD curves are indifferent to different amounts of data (aside from smoothness due to random variation), with performance remaining stable (relative to different constraints).

\subsection{Compression by updating the library}
A key assumption of the information-theoretic notion of compression that deviates from human learning is the invariance of the compression model throughout the learning process. 
One way of enabling the accumulation of knowledge from past experiences is to adapt the language (i.e., compression model) to observations through the ability to define new primitives. These new primitives are composed of subprograms that have been used successfully in previous tasks and cached in a library that is shared across melodies. 

For this goal, we use \textit{Adaptor Grammars} \citep[AGs;][]{johnson2006adaptor}, which augment PCFGs with a capacity for building a shared library of primitives, thus breaking the strong assumptions of independence in PCFGs. Specifically, we use an AG based on the Pitman-Yor process \citep{teh2006hierarchical}, as defined in \cite{liang2010learning}.  The AG model learns a distribution over grammars during training, and stores programs for future use in its cached library~$\mathcal{L}^{(i)}$ given all tasks~$X^{\{1:i\}}$ seen so far, updating the prior $(\pi|\mathcal{L}^{(i)})$ for future observations. 

Formally, for a collection of programs~$C_t$ of type~$t$ in the library, AGs construct new programs with probability~$\lambda_1$ and otherwise, it returns a cached program of type~$t$ with probability $\lambda_2$. 
let~$N_t$ be the number of distinct elements in~$C_t$, and~$M_\pi$ be the number of times~$\pi$ occurs in~$C_t$:
\begin{align}
    \lambda_1=\frac{\alpha+N_t d}{\alpha+\left|C_t\right|}, \quad \lambda_2=\frac{M_\pi-d}{\left|C_t\right|-N_t d}
\end{align}
Hyperparameters~$\alpha>0$ and~$0<d<1$ govern the degree of sharing and reuse, respectively. %
Smaller values for~$\alpha$ and $d$ lead to increased sharing and reduced construction, as $\lambda_1$ is proportional to $\alpha+N_t d$. Similarly, $\lambda_2$ is proportional to $M_\pi$, indicating that more frequently cached programs receive higher weights, irrespective of their internal complexity.

\paragraph{Computational efficiency.}
We expect the augmentation of PCFGs with a library to improve the efficiency of program induction in terms of both search budget and description length. For the search process, the iteratively refined prior $p(\pi|\mathcal{L})$ serves as the proposal distribution, leading to fewer rejected samples in the MC algorithm as more data is acquired.  
For description length, the introduction of new primitives can facilitate more compact representations, analogously to how defining new functions that can be called later enables a programmer to reduce code length.

Compared to PFCG, AG achieves lower distortion with the same resources, as well as being less sensitive to the search budget (Fig.~\ref{fig:rdt-result}c; note the different range of the x-axes). Furthermore, as the AG model acquires more data (Fig.~\ref{fig:rdt-result}d), the new primitives added to the library facilitate even more efficient and compact representations.

\begin{figure}
\centering
  \includegraphics[height=2.8cm]{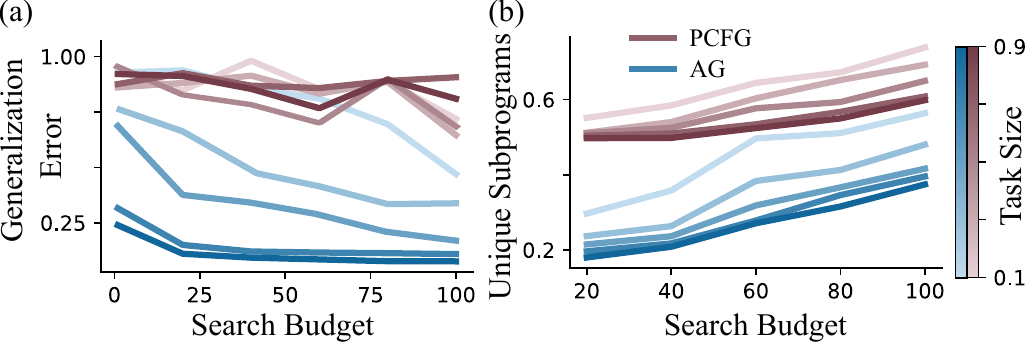}
  \caption{(a) Generalization performance given different search budgets. (b) The ratio of unique subprograms used in compressing different melodies. }
  \label{fig:ag-result}
  \vspace{-1.5em}
\end{figure}

To explore the intuition that an adaptive library leads to a more efficient search, we test one-shot generalization, where for a new, unobserved melody, a single program is sampled and evaluated. 
Figure~\ref{fig:ag-result}a shows the results under different search budgets and exposure to other melodies from the training data. Although both models perform similarly with very little training, generalization error for PCFG is largely unaffected by more training and a greater search budget. In contrast, the AG model improves substantially with both factors. 

We then analyze the degree of unique subprograms across melodies when increasing the search budget (Fig.~\ref{fig:ag-result}b). The results show that a greater search budget increases uniqueness by developing melody-specific representations. However, greater exposure to training data decreases uniqueness, by encouraging more reuse of subprograms across different melodies. Across both patterns, AG has substantially more shared subprograms during training (50\%-80\%) than PCFG (20\%-50\%). These results suggest that the greater generalization accuracy of AG is due to developing a more domain-adapted language.

\subsection{Curriculum effects}
We have demonstrated that augmenting the learner with a library enables it to adapt its language to the observed data, making the AG model more efficient for both description length and search constraints. However, this comes at the cost of sensitivity to curriculum effects. Consider that the augmented learning process is performing joint inference over both programs and the library $p(\pi,\mathcal{L}\,|\,X^{\{1:i\}})$
given all tasks~$X^{\{1:i\}}$ seen so far (Fig.~\ref{fig:framework}b). Ideally, the learner would maintain a full posterior over possible libraries $p(\mathcal{L}\,|\,X^{\{1:i\}})$. However, due to the combinatorial explosion of the hypothesis space, this is not feasible. Typically the posterior is approximated with a single point estimate \citep[or particle;][]{ellis2021dreamcoder, zhao2023model}. Without access to all previously encountered tasks, this is no longer a sufficient statistic for the past. Thus, there is sensitivity to the order in which training samples are encountered \citep{Nagy2016, zhao2023model, dekker2022curriculum}. 

\begin{figure}
\centering
  \includegraphics[height=4cm]{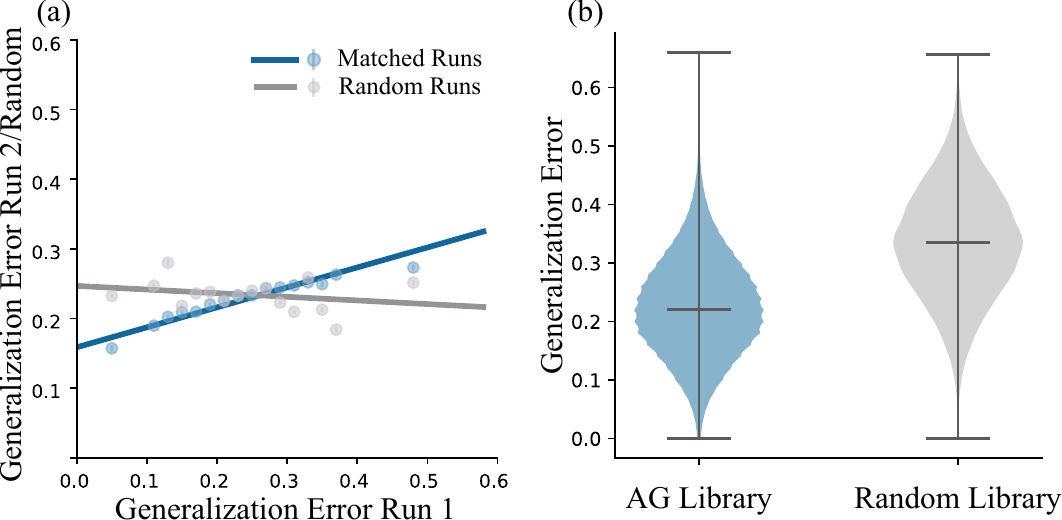}
  \caption{Curriculum effect in AGs. (a) Comparison of matched runs of the same curriculum vs. different curricula. (b) Generalization performance of  AGs with a curriculum-informed library vs. a randomized library. }
  \label{fig:ag-curriculum}
  \vspace{-1em}
\end{figure}

Due to factorial growth in the number of unique stimuli orderings (i.e., curricula) as a function of the number of melodies, we ran simulations with 50 melodies across 1,000 randomly sampled curricula.
Empirically, we observe high variability of learned libraries across curricula, with an average of only $3.19\% \pm 0.05\%$ shared subprograms under different curricula.
To disentangle curriculum effects from the inherent stochasticity of the sampling during inference, we ran the model twice for curricula. We then compared the generalization error between matched runs of the same curriculum, compared against a random baseline with two different curricula (Fig.~\ref{fig:ag-curriculum}a). 
Matched runs were positively correlated ($r(998)=0.286$, $p<.001$), whereas random runs were somewhat negatively correlated ($r(998)=-0.052$, $p<.001$), with the difference in performance significantly larger in matched vs. random runs ($t(999)=7.621$, $p<.001$).
These results demonstrate that the curriculum effect cannot be accounted for by stochasticity alone.

Having established that learned libraries are influenced by different curricula, we compare generalization performance between AGs' curriculum-dependent libraries against randomly generated libraries over 1,000 simulations (Fig.~\ref{fig:ag-curriculum}b). The random libraries consisted of randomly sampled subprograms, which are used to compress the same set of tasks.
While random libraries perform worse than AG libraries on average in generalizing to new melodies, the variability of performance is quite similar, meaning unfavorable learning orders may potentially disadvantage AGs in some settings. %

\subsection{Synergistic curricula} 
While AGs have been used to explain curriculum effects in human learning \citep{zhao2023model}, previous research has typically used simple tasks with orthogonal features \citep{dekker2022curriculum, rule2018learning}, making curriculum design trivially easy.
Here, we explore whether we can \textit{a priori} generate beneficial curricula for generalizing to new melodies.

\begin{figure}
\centering
  \includegraphics[height=3.3cm]{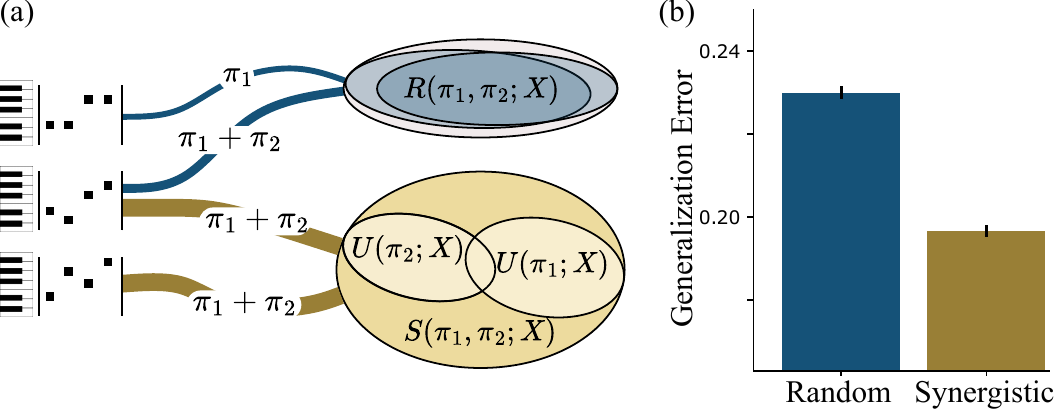} 
  \caption{(a) Example libraries learned by AGs with random curricula (blue) and learned with synergistic curricula (yellow; including $R\textrm{edundant}$, $U\textrm{nique}$, and $S\textrm{ynergistic}$ information) learned under different curricula. (b) Generalization performance given random and synergistic curricula. }
  \label{fig:synergy}
\end{figure}

We introduce a synergistic curriculum building method using principles from partial information decomposition \citep[PID;][]{williams2010nonnegative, proca2022synergistic, luppi2024information}. 
PID decomposes multivariate mutual information~$I(\pi_1, \ldots, \pi_M; X)$ between multiple sources (i.e., the programs~$\pi_m \in \mathcal{L}$) and a target variable (i.e., melodies~$X$) into different components (Fig.~\ref{fig:synergy}a): \emph{unique}~$U(\pi_m; X)$ information present in exactly one program~$\pi_m$, \emph{redundant}~$R(\pi_i, \ldots, \pi_j; X)$ information encoded by separately in multiple programs, and \emph{synergistic} information~$S(\pi_1, \ldots, \pi_M; X)$ jointly encoded by multiple programs. If the library~$\mathcal{L}$ contains two programs, the synergy can be represented as 
\begin{align}
   S(\mathcal{L}; X) =  I(\pi_1, \pi_2 ; X)-R(\pi_1, \pi_2 ; X)-U(\pi_1 ; X)-U(\pi_2 ; X)
\end{align}
Synergy and related quantities, such as integrated information, have been found in complex information processing \citep{mediano2021towards}. We hypothesize that for both artificial and biological agents, a synergistic library is essential for being able to flexibly learn and generalize across novel problems.
Here, we test whether synergy can guide the design of curricula for our AG model, with potential future applications for human learners \citep{zhou2024predictive}.
Specifically, we strategically select and order melodies with the goal of maximizing synergy. Formally, given the learned library~$\mathcal{L}^{(n)}$ at $n$-th step, we select the next presented melody~$X^*$ by maximizing the synergy of the potential library~$S(\mathcal{L}; X)$
\begin{align}
    X^* = \mathrm{argmax}_{X^{(i)} \in X}\, S(\mathcal{L}; X) \quad \text{where} \,\, \mathcal{L} \sim p(\mathcal{L}^{(n+1)} \,|\, \mathcal{L}^{(n)}, X^{(n)})
\end{align}
The aim is to construct a sequence of training data, where each new melody is expected to add subprograms to the library that enhances the overall representational capacity.
To test this method, we show that generalization performance under a synergistic curriculum is better than a random ordering ($t(999)=6.219$, $p<.001$; Fig.~\ref{fig:synergy}b).

\section{Discussion}
In this work, we integrated \textit{Rate Distortion Theory} (RDT) with \textit{Bayesian program induction} to explore how the dual constraints of description length and search complexity influence representation learning. Our simulations demonstrate that creating a shared program library across tasks facilitates more efficient learning and generalization, relative to both constraints. We also demonstrated how partial information decomposition provides an \textit{a priori} method to design curricula that foster the development of more effective program libraries in learners, with potential future applications for human learners.

We have only begun to explore the hypothesis space in this domain, and there is still a large gap between our simulations and being able to model human learners with similar models. Nevertheless, this work presents one of the first attempts to integrate two important principles of human learning and has shed light on new patterns that were not obvious from existing theories. 
An important limitation that needs to be addressed for modeling human learning is that the greedy segmentation of melodies and subsequent independent generation of programs does not allow for capturing long-range structure within-melodies. 
This segmentation also makes targeting a specific rate difficult. While we expect that the term deletion process we introduced to enable fine-grained control over the rate might have cognitive plausibility, it may overestimate the distortion for some resource constraints. 
Furthermore, when defining new primitives, our model only considers the current task, and does not consider refactored versions of subprograms \citep{ellis2021dreamcoder}, therefore missing chances for further sharing of components. 
Finally, in future experiments, we aim to explore how different design choices, such as alternative distortion functions, prior distributions, and task domains beyond melodies, would impact our results.

Overall, the integration between compression and program learning is a very promising avenue for future research, and here, we have taken the first steps in this direction.

\clearpage
\section{Acknowledgments}
We are grateful to Peter Dayan for helpful comments.
The authors thank the International Max Planck Research School for Intelligent Systems (IMPRS-IS) for supporting Hanqi Zhou. 
This research was supported as part of the LEAD Graduate School \& Research Network, which is funded by the Ministry of Science, Research and the Arts of the state of Baden-Württemberg within the framework of the sustainability funding for the projects of the Excellence Initiative II.
This work is supported by the German Federal Ministry of Education and Research (BMBF): Tübingen AI Center, FKZ: 01IS18039A, funded by the Deutsche Forschungsgemeinschaft (DFG, German Research Foundation) under Germany’s Excellence Strategy–EXC2064/1–390727645, and funded by the DFG under Germany's Excellence Strategy – EXC 2117 – 422037984.

\bibliographystyle{apacite}

\setlength{\bibleftmargin}{.125in}
\setlength{\bibindent}{-\bibleftmargin}

\end{document}